\documentclass[aps,prl,twocolumn,amsmath,amssymb,superscriptaddress]{revtex4}

\usepackage{graphicx}
\usepackage{color}

\begin{document}

\preprint{preprint(\today)}

\title{Uniaxial Strain Tuning of Superconductivity in 2$H$-NbSe$_{2}$}

\author{Andrew Wieteska}
\email{jrw2200@columbia.edu} 
\affiliation{Department of Physics, Columbia University, New York, NY 10027, USA}

\author{Ben Foutty}
\affiliation{Department of Physics, Columbia University, New York, NY 10027, USA}

\author{Zurab Guguchia}
\affiliation{Department of Physics, Columbia University, New York, NY 10027, USA}
\affiliation{Laboratory for Muon Spin Spectroscopy, Paul Scherrer Institute, CH-5232 Villigen PSI, Switzerland}

\author{Felix Flicker}
\affiliation{Rudolph Peierls Centre for Theoretical Physics, University of Oxford, Department of Physics, Clarendon Laboratory, Parks Road, Oxford, OX1 3PU, United Kingdom}

\author{Ben Mazel}
\affiliation{Department of Physics, Columbia University, New York, NY 10027, USA}

\author{Lyuwen Fu}
\affiliation{Department of Applied Physics and Applied Mathematics, Columbia University, New York, NY 10027, USA}

\author{Shuang Jia} 
\affiliation{International Center for Quantum Materials, Peking University, Beijing 100871, China}

\author{Chris Marianetti}
\affiliation{Department of Applied Physics and Applied Mathematics, Columbia University, New York, NY 10027, USA}

\author{Jasper van Wezel}
\affiliation{Institute for Theoretical Physics, Institute of Physics, University of Amsterdam, 1090 GL Amsterdam, The Netherlands}

\author{Abhay Pasupathy}
\email{apn2108@columbia.edu} 
\affiliation{Department of Physics, Columbia University, New York, NY 10027, USA}

\begin{abstract}
We explore the effect of lattice anisotropy on the charge-ordered superconductor 2H-NbSe$_{2}$. Using a novel strain apparatus, we measure the superconducting transition temperature $T_{sc}$ as a function of uniaxial strain. It is found that $T_{sc}$ is independent of tensile(compressive) strain below a threshold of 0.2\% (0.1\%), but decreases strongly with larger strains with an average rate of $1.3\,$K/\% ($2.5\,$K/\%). Transport signatures of charge order are largely unaffected as a function of strain. Theoretical considerations show that the change in the behavior of $T_{sc}$ with strain coincides with a phase transition from 3Q to 1Q charge order in the material. The spectral weight on one of the Fermi surface bands is found to change strongly as a consequence of this phase transition, providing a pathway to tune superconducting order. 
\end{abstract}

\maketitle

\begin{figure*}[t!]
\centering
\includegraphics[width=1.0\linewidth]{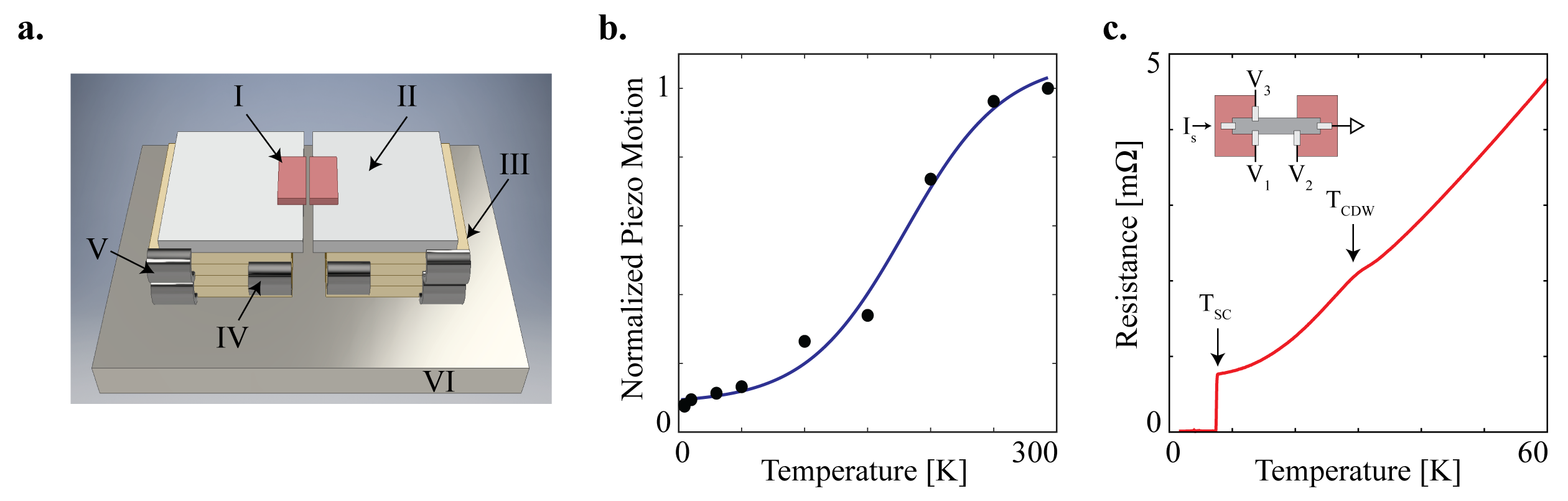}
\caption{(Color online) (a) Schematic illustration of the uniaxial strain device. A pair of stacks of shear piezoelectric actuators (III) are mounted on a titanium plate (VI). A sample is mounted on rectangular ruby plates (I) which in turn are mounted on sapphire caps (II) that insulate the sample from the electrical contacts to the piezos. Voltage applied across contacts (IV-V) continuously tunes the spacing between the stacks, and the corresponding strain applied to the sample. (b) Measured temperature dependence of the relative displacement of PbZr shear piezoelectric actuators, normalized to motion at 300 K. (c) Temperature-dependence of resistivity of NbSe$_{2}$. Inset shows the 5-probe contact geometry (see main text for details).} 
\label{fig1}
\end{figure*}

Unconventional superconductivity is often found in the proximity of symmetry-breaking phases. In cuprate and pnictide high-$T_c$ superconductors there exist multiple phases that break translational \cite{tranquada, CDWcuprates,CDWybco,pnictides_stripes_neutrons} and rotational \cite{fisher_anisotropy,ybco_shg,bozovic_films} symmetries. Furthermore, rotational symmetry breaking has been reported in the superconducting state of intercalated Bi$_2$Se$_3$ \cite{bi2se3_nmr,bi2se3_maeno}, in the heavy fermion superconductor URu$_2$Si$_2$ \cite{uru2si2_fisher} and in the recently discovered twisted magic angle bilayer graphene superconductor \cite{alex_magic_angle,pablo_magic_angle_nematic}. The broad question of interplay between symmetry breaking and superconductivity remains the subject of intense debate \cite{kivelsonkeimer}. 

Experimentally, application of symmetry-breaking strains to unconventional superconductors has been a fruitful line of investigation. In the iron-pnictides, anisotropic biaxial strain couples directly to the order parameter of the nematic transition \cite{fisher_ironbased_strain} and, for high strains, acts as a tuning parameter for the transition \cite{fisher_ironbased_symm_antisymm}. A recent inelastic X-ray scattering study on cuprates showed that a novel charge-ordered phase forms when samples are put under large uniaxial pressures \cite{cuprates_rixs_strain}. In Sr$_2$RuO$_4$, the effects of anisotropic strain on $C_4$ symmetric superconducting phased are spectacular \cite{ruthenate214_lowstrains,ruthenate214_highstrains}.

The effect of anisotropic strain on the charge ordered superconductor 2H-NbSe$_2$ is the subject of this letter. NbSe$_2$ is a BCS, two-gap $s$-wave superconductor with $T_{sc}=7.2\,$K in which a charge density wave distortion onsets below $T_{cdw}=33.7\,$K \cite{NbSe2CDWdiscovery, TakagiARPES}. This (`3Q') charge ordered phase possesses $C_3$ rotational symmetry and its period is $\sim 3 \times 3$ lattice constants. The origin of the distortion seems to be strong electron-phonon coupling rather than by a Fermi surface instability \cite{diSalvo1975, NbSe2basov, WeberEA11, abhaypaper1, abhaypaper2, FF_JvW_NatComms15}. Two recent scanning tunnelling microscopy studies \cite{hoffmanstm, FF_JvW_PNAS18} reported evidence of a novel, unidirectional charge ordering in samples where strain is present. In the current work we measure the change in the superconducting critical temperature $T_{sc}$ in NbSe$_2$ under large, continuously tunable uniaxial strain. Our main result is to reveal a large, negative uniaxial strain effect on $T_{sc}$ in highly strained samples. These data can be interpreted in terms of differences in Fermi surface competition between superconductivity and charge order in the $C_3$-symmetric and $C_3$ symmetry-breaking charge ordered phases.
 
Electrical transport measurements were performed on high-quality single crystals of 2H-NbSe$_2$. Uniaxial strain was applied using a piezoelectric strain application device shown in Figure 1a. The device consists of two stacks of piezoelectric shear actuators across which a sample is mounted. Strain is applied \emph{in situ} by tuning the separation between the stacks with external voltage. The temperature dependence of the motion of the piezoelectric stacks was calibrated using a home-built capacitative sensor, as shown in Figure 1b. Differential thermal contraction between components of the device gives a correction to applied strain (estimated to be about 0.08$\%$ tension at superconducting temperatures), but this correction is approximately temperature independent in the range of our measurement (below $85\,$K). NbSe$_2$ crystals were mounted with the $c$-axis perpendicular to the direction of strain application. Longitudinal resistivity of the strained crystal was measured in the standard four-probe geometry, with $R_{xx}$ measured across the strained region of the crystal (see Fig. 1a inset). Hall resistivities were extracted through antisymmetrization of magnetoresistivity measured for the two out-of-plane field directions.  A representative resistivity curve for our NbSe$_2$ samples is shown in Fig. 1c. Clearly observable are an anomaly around $T_{cdw}=33.7\,$K and a sharp superconducting transition at $T_{sc}=7.2\,$K. 

Figures 2a and 2b show the temperature dependence of the resistivity for NbSe$_2$, for a range of tensile and compressive strains. The superconducting transition is extremely sharp ($<0.1\,$K width) with no applied strain and broadens only moderately upon application of the highest strains we achieved ($<0.2\,$K width). At low strains, below 0.2\% tension and 0.1\% compression, we did not find a measurable change in $T_{sc}$. At higher strains, $T_{sc}$ is depressed until a minimum of $6.5\,$K ($5.5\,$K) for 0.55\% extension (0.6\% compression). We did not observe measurable elastoresistance in the normal state up to the highest strains we applied. The experiment was repeated for multiple samples (one additional set of data from Sample B is shown in Fig.~3) and a consistently large strain dependence of $T_{sc}$ was seen.  

\begin{figure}[b!]
\centering
\includegraphics[width=\linewidth, scale=1]{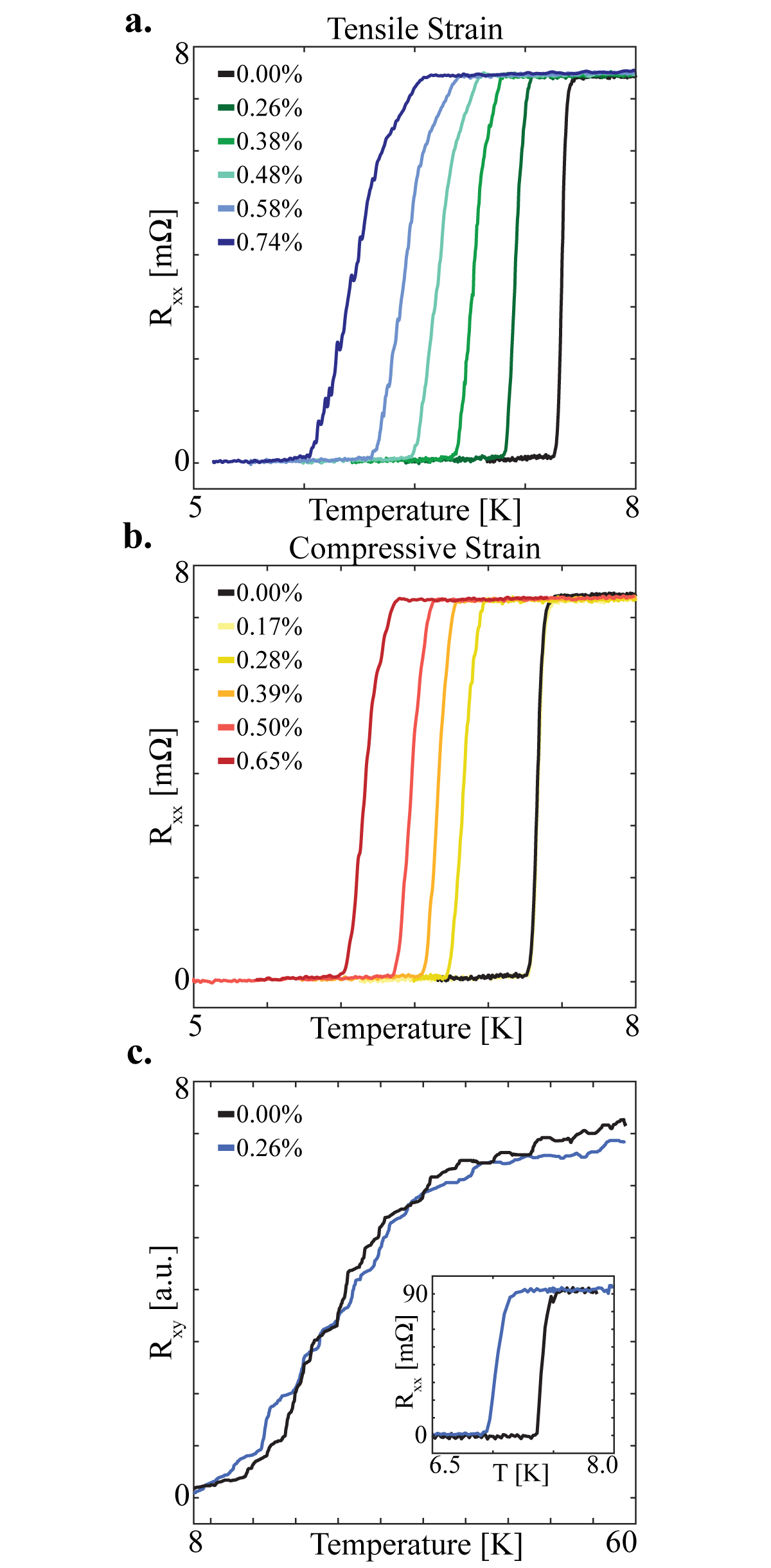}
\caption{ (Color online) (a) The superconducting transition of NbSe$_2$ at different tensile (a) and compressive (b) strains. c) Temperature-dependent Hall resistance at two different strains. The inset shows the superconducting transitions corresponding to the two Hall curves. }
\label{fig3}
\end{figure}

A natural follow-up question is how $T_{cdw}$ depends on uniaxial strain. Previous works have observed a small signature of the charge density wave transition in the longitudinal resistivity at $T_{cdw}$ in the cleanest samples. This feature is highly dependent on the disorder level of the samples, since in the presence of disorder the CDW transition is significantly smeared out \cite{NbSe2disorder}. In our samples, this anomaly in the longitudinal resistivity is extremely weak. We were not able to reliably measure changes in this feature as a function of strain. An alternative signature in transport is provided by the Hall coefficient, which has been shown to have a sign change at a temperature a little below the CDW transition. Figure 2c shows the temperature dependence of the Hall resistance for sample A, recorded at zero strain and under the applied tensile strain of 0.4\%. We identify $T_{cdw}$ with the onset temperature of a downturn in the Hall resistance, which falls around $T=35\,$K. Our measurement shows that the CDW transition temperature does not shift within this strain interval by more than the sensitivity of this measurement, estimated to be $\Delta T\sim 2\,$K. We note that for the same change in strain, there is a significant drop in $T_{sc}$ (see inset of Fig.~2c). Consequently, the effect of uniaxial, $ab$-plane strain on $T_{cdw}$ in this material is weak compared to the effect on $T_{sc}$. Precise determination of how strain affects $T_{cdw}$ will require combining uniaxial strain with a direct probe of charge order, such as high-resolution X-ray diffraction or Raman spectroscopy.

\begin{figure}[b!]
\centering
\includegraphics[width=\linewidth, scale=1]{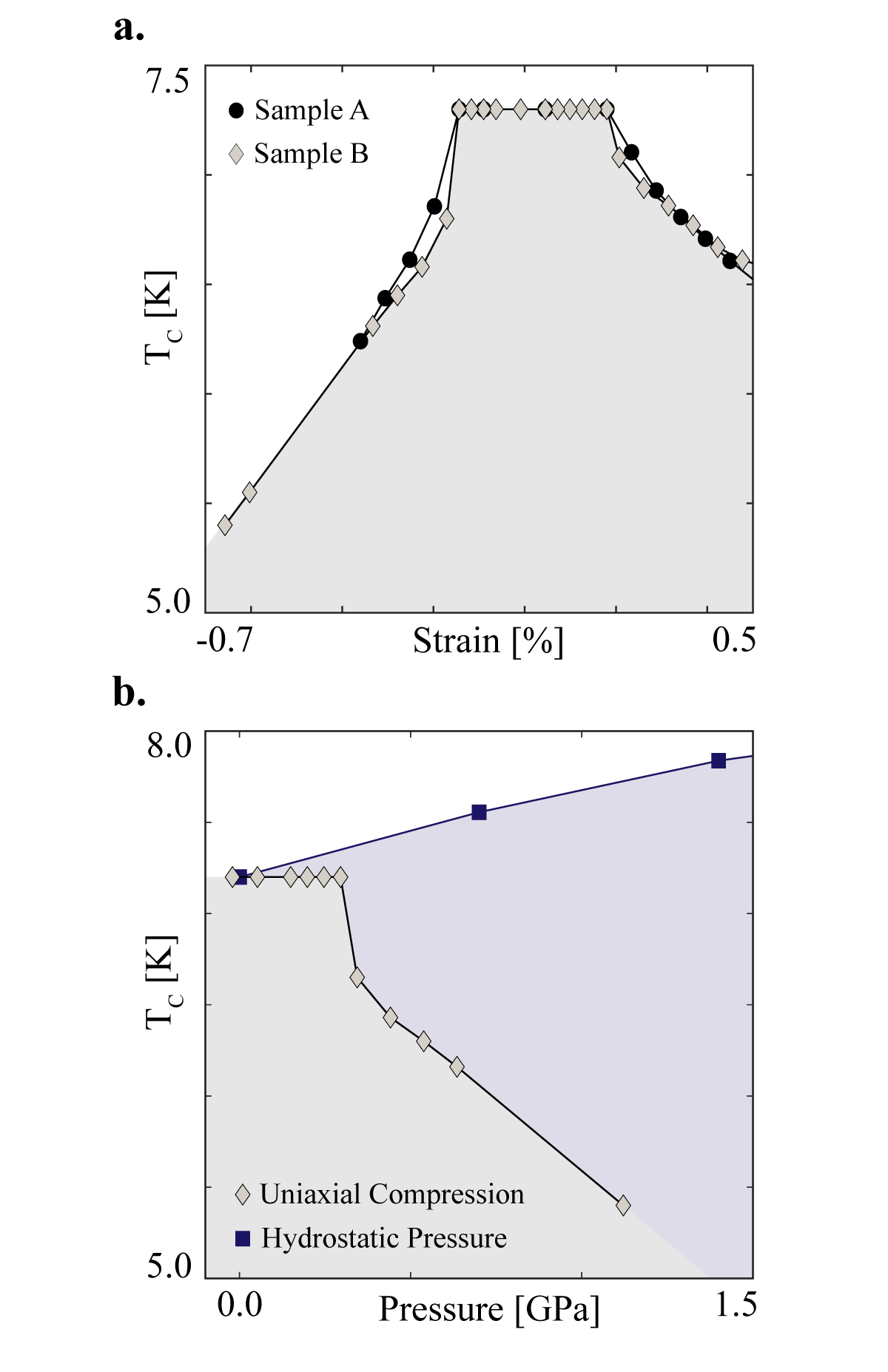}
\caption{ (Color online) (a) Strain dependence of superconducting $T_{sc}$ for two different samples. (b) Comparison between hydrostatic and uniaxial pressure effects on superconducting $T_{sc}$.}
\label{fig3}
\end{figure}
 
The results are summarized in Fig.~3a in the form of a phase diagram showing the dependence of $T_{sc}$ on the tensile and compressive strains. The results are as follows: 1) $T_{sc}$ is insensitive to strain within a region centered around zero strain; 2) beyond this region, $T_{sc}$ is significantly suppressed with both tension and compression; 3) compressive strain has a stronger effect on $T_{sc}$ than tensile strain, by about a factor of two. We contrast these results of measurements as function of uniaxial strain with the hydrostatic pressure dependence of $T_{sc}$. We have considered the possibility that small strains are not properly transmitted to the sample. By measuring the elastoresistance of well-known iron pnictides \cite{fisher_ironbased_strain} at small strains, we have confirmed that strains down to <0.01$\%$ are transmitted well to the sample.  In Fig.~3b we summarize the hydrostatic pressure data based on the previously known pressure dependence of $T_{sc}$ \cite{pressureprb, pressurerosenbaum}. Interestingly, the sign of the effect is opposite for uniaxial and hydrostatic pressure: hydrostatic pressure monotonically increases $T_{sc}$ whereas the strain dependence of $T_{sc}$ contains a transition between a region of no dependence to a region where $T_{sc}$ is significantly affected. The magnitude of the uniaxial strain effect is much larger ($1.8\,$K at 0.6\%) than the corresponding change under hydrostatic pressure ($0.6\,$K). 

We now consider possible theoretical explanations for our data. Within electron-phonon coupling-based BCS mechanisms for superconductivity, there are a few possible explanations for a change in $T_{sc}$: changes in the number of states available for pairing at the Fermi level, changes in the phonon dispersion and in the electron-phonon coupling, as well as changes in electron-electron interactions (screening). We first ignore the presence of the CDW in the material, and ask whether the changes in these parameters that are caused by uniaxial strain are sufficient to explain the change in $T_{sc}$ observed experimentally. We investigated this using density functional theory (DFT) calculations within the Generalized Gradient Approximation (GGA)\cite{Perdew19926671}. These calculations were performed using the
Projector Augmented Wave (PAW) method \cite{Blochl199417953,Kresse19991758}, as implemented in the VASP code \cite{Kresse1993558,Kresse199414251,Kresse199615,Kresse199611169}.  A plane wave basis with a kinetic energy cutoff of $550\,$eV was employed. We used a $\Gamma$-centered \textbf{k}-point  mesh of 20$\times$20$\times$10. The crystal structure was relaxed, yielding lattice parameters of $\vec a_1=a/2(\sqrt 3,\bar 1,0)$, $\vec a_2=a/2(\sqrt 3,1,0)$, and $\vec a_3=(0,0,c)$, where $a=3.5038\,$\AA, and $c=13.7213\,$\AA; while the distance between the Nb and Se layers was $1.6734\,$\AA. The distance measure for all band structure plots is in lattice coordinates, such that different strains can easily be compared. The net result of these calculations shows that the basic electron and phonon structure do not change with strain in a fashion that can explain the observed changes in $T_{sc}$, especially the nonlinear behavior of $T_{sc}$ as a function of strain.

\begin{figure*}[t!]
\centering
\includegraphics[width=1.0\linewidth]{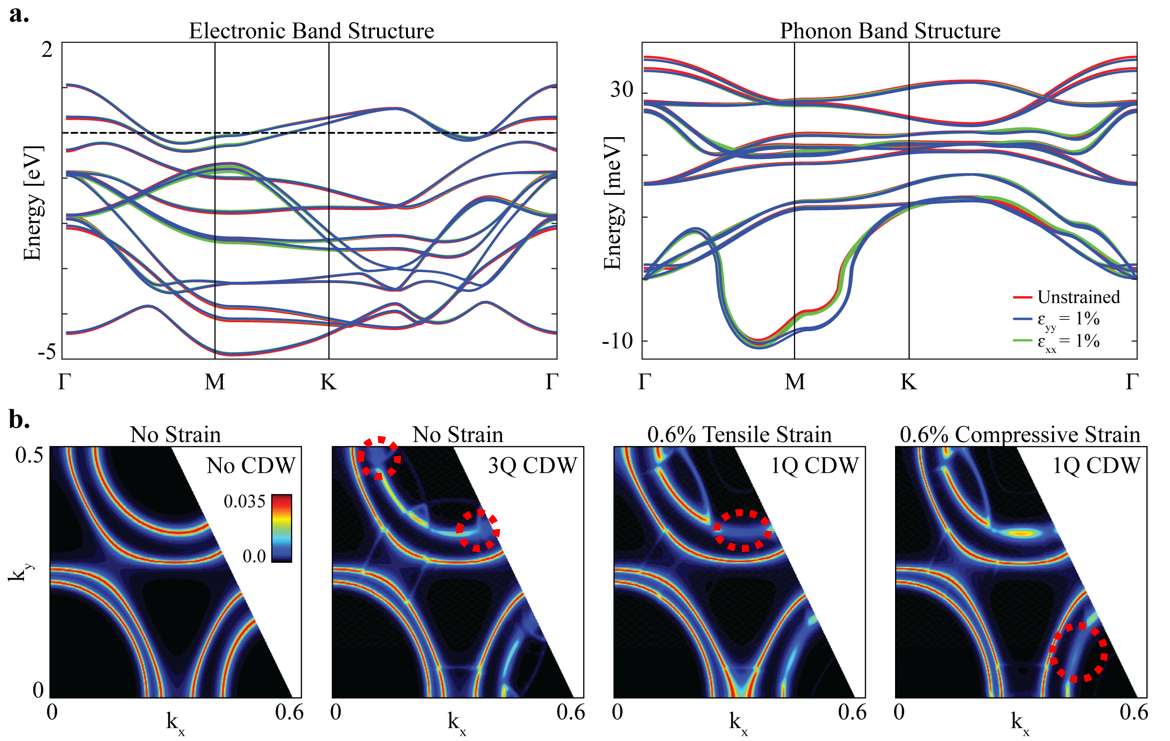}
%
\caption{(Color online) (a) Electronic band structure and phonon spectra computed from DFT for different uniaxial strains. (b) ARPES spectral density at the Fermi level for the unstrained and undistorted material (left panel), the unstrained and `3Q' distorted material (second panel from the left), $0.6\%$ extension and `1Q' distorted (second panel from the right) and $0.6\%$ compression and `1Q' distorted (right panel). Regions where the spectral density is suppressed due to charge ordering are highlighted with red circles.
} 
\label{fig4}
\end{figure*}

We next consider possible changes to the CDW structure as a function of strain, and how such changes could affect $T_{sc}$. To do so, we describe charge ordering in NbSe$_2$ within the random phase approximation (RPA), that takes into account the momentum dependence of the electron-phonon coupling as well as the orbital content of the different bands. This model has been shown to provide a consistent explanation of the full range of experimental results in NbSe$_2$~\cite{FF_JvW_NatComms15,FF_JvW_PRB15,FF_JvW_PRB16}, including the impact of strain on charge ordering geometries~\cite{FF_JvW_PNAS18}. Our prior calculations suggest that a transition from the 3Q to the 1Q phase occurs with increasing strain~\cite{FF_JvW_PRB15,FF_JvW_PRB16} and we address the effect of each of these CDW phases on the superconducting order in the present work. The simplest way in which CDW order affects superconductivity is by gapping out parts of the Fermi surface and thus lowering $T_{sc}$. We test this possibility through a calculation of the spectral function $A\left(E_F,\mathbf{k}\right)$ in each of the two CDW phases at various strains~\cite{NormanEA99}. Our results are shown in Fig.~4b for the four relevant cases: zero strain, with no CDW gap (4a); zero strain, 3Q CDW gap ; $0.6\%$ compressive strain and $0.6\%$ tensile strain (4d) with 1Q CDW gap.

To understand these results, we recall that the Fermi surface (FS) consists of quasi-two-dimensional double-walled cylinders centered around the $K$ point, and a three-dimensional pancake band centered around the $\Gamma$ point. ARPES measurements have shown that the 3Q CDW gap opens only on a small area of the FS and largely on the inner pocket centered around the $K$ point (dashed red circles, figure 4b, second panel). Further, low temperature ARPES measurements also place the dominant superconducting gap on that pocket. Turning to our data, we note that in the 1Q phase the Brillouin zone contains two, now inequivalent, inner K-pockets and that the charge order gap opens on much larger sections of these pockets in both of the strain geometries. Since these pockets are where the superconducting gap resides in the unstrained material, this confirms the idea that the suppression of $T_{sc}$ at higher strains is due to competition with the CDW. Significantly, the two-fold asymmetry in magnitudes of the slopes of transition lines for tensile and compressive strains follows naturally from the model upon assuming the unit cell area is approximately conserved~\cite{FF_JvW_PRB15,FF_JvW_PRB16}.

In conclusion, we have studied the effect of large, tunable uniaxial strain on $2H$-NbSe$_2$ single crystals. We found that the superconducting transition temperature $T_{sc}$ is insensitive to uniaxial strains less than 0.2\% (0.1\%) tension (compression), but that at higher strains $T_{sc}$ is significantly depressed, by up to 25\%. We interpret our data in terms of a strain-driven transition between triangular (3Q) and unidirectional (1Q) charge ordering geometries and competition for spectral weight between the 1Q CDW and superconductivity at high strains. Future studies of uniaxial strain effects in this important model system could focus on obtaining the precise strain dependence of the charge ordering using X-ray diffraction and of the anisotropies in the superconducting order parameter using probes such as muon spin rotation and scanning tunneling spectroscopy.

\textbf{\section{Acknowledgments}}

We thank D. Shahar and E. Telford for experimental help. This work is supported by the National Science Foundation via grant DMR-1610110. Equipment support is provided by the Air Force Office of Scientific Research via grant FA9550-16-1-0601. FF acknowledges support from the Astor Junior Research Fellowship of New College, Oxford. JvW acknowledges support from a VIDI grant financed by the Netherlands Organization for Scientific Research (NWO). CAM and LF were supported by the grant DE-SC0016507 funded by the U.S. Department of Energy, Office of Science. This research used resources of the National Energy Research Scientific Computing Center, a DOE Office of Science User Facility supported by the Office of Science of the U.S. Department of Energy under Contract No. DE-AC02-05CH11231.

\end{document}